\documentclass{article}[11pt]

\usepackage[a4paper,top=4.0cm,bottom=2.0cm,left=2.0cm,right=2.0cm]{geometry}

\usepackage{amsmath,amsfonts,mathrsfs,amsfonts,ulem}
\usepackage{epsfig,graphicx}
\usepackage{color}
\usepackage{todonotes}
\usepackage{enumerate}
\usepackage{lineno}

\usepackage{amsmath,amsfonts,mathrsfs,amsfonts, color, bbold,bm,blindtext}

\usepackage{algorithmicx,algorithm,algpseudocode}

\allowdisplaybreaks

\usepackage{textcomp}

\begin{document}

\title{Thermophysical modelling and parameter estimation of small solar system bodies via data assimilation}

\author{M. Hamm\thanks{Universit\"at Potsdam, 
			Institut f\"ur Mathematik, Karl-Liebknecht-Str. 24/25, D-14476 Potsdam, Germany ({\tt maximilian.hamm@dlr.de}) and German Aerospace Center (DLR), Rutherfordstr. 2, 12489 Berlin,} \and I. Pelivan\thanks{Fraunhofer Heinrich Hertz Institute (HHI), Einsteinufer 37, 10587 Berlin, Germany ({\tt ipelivan@gmx.net})
} \and M. Grott\thanks{German Aerospace Center (DLR), Rutherfordstr. 2, 12489 Berlin, Germany ({\tt matthias.grott@dlr.de})
} \and J. de Wiljes\thanks{Institut f\"ur Mathematik, Karl-Liebknecht-Str. 24/25, D-14476 Potsdam, Germany ({\tt wiljes@uni-potsdam.de})
}}
\maketitle
\begin{abstract}

Deriving thermophysical properties such as thermal inertia from thermal infrared observations provides useful insights into the structure of the surface material on planetary bodies. The estimation of these properties is usually done by fitting temperature variations calculated by thermophysical models to infrared observations. For multiple free model parameters, traditional methods such as Least-Squares fitting or Markov-Chain Monte-Carlo methods become computationally too expensive. Consequently, the simultaneous estimation of several thermophysical parameters together with their corresponding uncertainties and correlations is often not computationally feasible and the analysis is usually reduced to fitting one or two parameters.
Data assimilation methods have been shown to be robust while sufficiently accurate and computationally affordable even for a large number of parameters.  This paper will introduce a standard sequential data assimilation method, the Ensemble Square Root Filter, to thermophysical modelling of asteroid surfaces. This method is used to re-analyse infrared observations of the MARA instrument, which measured the diurnal temperature variation of a single boulder on the surface of near-Earth asteroid (162173) Ryugu. The thermal inertia is estimated to be $295 \pm 18$ $\mathrm{J\,m^{-2}\,K^{-1}\,s^{-1/2}}$, while all five free parameters of the initial analysis are varied and estimated simultaneously. Based on this thermal inertia estimate the thermal conductivity of the boulder is estimated to be between 0.07 and 0.12 $\mathrm{W\,m^{-1}\,K^{-1}}$ and the porosity to be between 0.30 and 0.52. For the first time in thermophysical parameter derivation, correlations and uncertainties of all free model parameters are incorporated in the estimation procedure which is more than 5000 times more efficient than a comparable parameter sweep.
\end{abstract}



\section{Introduction}

Thermal conditions on atmosphereless, small solar system bodies are governed by the thermophysical properties of the surface material, e.g. thermal conductivity, heat capacity, and emissivity. The thermal conductivity is coupled to structural properties of the surface material such as grain size and porosity \cite{sakatani2017}. Observing the surface in the thermal infrared wavelength range, typically 5-25 \textmu{}m, provides direct insight into the thermal conditions on the surface. Thus, thermophysical and structural material properties can be derived from thermal infrared data.

The thermophysical properties of numerous solar system bodies have been investigated using telescopes \cite{Masiero2011,Harris2016,mueller2014,mueller2017}, or satellite remote sensing data \cite{Chase1969,Kieffer1972,Kuehrt1992,Christensen2001,Mellon2000,Fergason2006a,Paige2010}, as well as close-up studies performed by rovers and landers \cite{Fergason2006b,Spohn2015,GomezElvira2012,Hamilton2014,Vasavada2017,grott2019}. Recently, the Japanese Hayabusa2 mission \cite{watanabe2017} investigated the C-type near-Earth asteroid (162173) Ryugu with four instruments, including a thermal infrared imaging system \cite{okada2017,watanabe2019,sugita2019,okada2020}. The mission included the MASCOT lander \cite{ho2017} that, among other instruments, carried a thermal infrared radiometer \cite{grott2017}. MASCOT landed on the surface of Ryugu and investigated a single boulder on the surface of Ryugu for 2.5 asteroid rotations \cite{jaumann2019,scholten2019descent,scholten2019boulder,preusker2019}, recording a full diurnal surface temperature curve \cite{grott2019}. The NASA OSIRIS-REx mission is currently investigating the B-type near-Earth asteroid (101955) Bennu \cite{lauretta2019} using the OTES instrument to investigate the thermal properties of Bennu's surface \cite{christensen2018,dellagiustina2019}. Earlier, the Rosetta mission, consisting of an orbiter and a lander module, arrived at comet 67P/Churyumov-Gerasimenko (67P) and studied the comet in detail which included measurements in the thermal infrared on the surface of 67P \cite{Spohn2015}. 

Infrared data is usually analysed by comparing the observed flux to the results of thermophysical models  \cite[e.g][]{hamm2018,Pelivan2017,2018MNRAS.478..386P}, to fit the observation in a weighted least-squares approach  \cite{Nowicki2007,Spohn2015,Vasavada2017,dellagiustina2019,grott2019}. Typically, only a few parameters are varied in these works while most parameters of the model are assumed to be some constant value, or varied in coarse steps.
This is due to the extensive computation time necessary to compute a solution of the standard thermophysical models. Often, look-up tables are computed prior to the fitting, and the temperature variation of the surface is interpolated from these tables \cite{Nowicki2007}.

Recently, \cite{cambioni2019} published an approach where a surrogate model, in form of a neural network, returns the temperature variation of an asteroid surface given insolation data and some thermal parameters, i.e., thermal inertia of a rock component of the surface regolith, thermal inertia of the fine components, a surface roughness parameter, and the rock component's area coverage. While significantly increasing the speed of the temperature calculation and thus allowing to use Markov Chain Monte-Carlo approaches (MCMC) to approximate unknown parameters, the trained network is merely a surrogate model of the true physical system and is thus limited in its predictive power. Furthermore, the computational complexity to fit a neural network significantly increases for more detailed thermal models with a higher number of free parameters.  

The commonly used least-squares approach does not require to generate as many model evaluations as is necessary for a Monte-Carlo estimation but demands some form of linearisation. Consequently, unlike the Monte-Carlo ansatz, the least-squares technique can only  provide a Gaussian approximate of the true uncertainty of the parameter estimate. The approach presented in this paper addresses the issues associated with existing fitting algorithms such as the least-squares approach and MCMC estimation. More precisely the proposed method is computationally feasible, i.e., only a relatively small number of samples compared to the MCMC approach are necessary for the algorithm to give robust results. Further the method does not require a linearisation and thus is able to capture the highly nonlinear relationship between surface temperature and observable infrared emission while providing a good representation of the uncertainty of the approximation.

It is important to mention that the method presented in this paper is a standard approach that has been developed in the field of data assimilation (DA) \cite{jdw:EvensenLeeuwen2000,jdw:stuart15}. Here it is adapted to thermophysical modelling for the purpose of retrieving thermophysical properties from infrared observations. The proposed method, the so called Ensemble Square Root Filter (ESRF) \cite{jdw:tippett03,jdw:reichcotter15,Nergeretal2012}, combines the key strength of the least squares method, more specifically the "best linear unbiased estimator" \cite{jdw:reichcotter15}, with the ones of the Monte Carlo approach and it has been successfully applied to highly nonlinear problems with large number of free parameters of order $10^7$ and its accuracy and stability has been rigorously investigated in recent years \cite{deWiljesTong2019,deWiljesStannatReich2019,LangeStannat2019}.

Data assimilation techniques are widely employed in the Earth sciences, in particular in meteorology, atmospheric physics and oceanography. For other solar system bodies, data assimilation has been applied to atmospheric data sets from orbital Mars missions \cite{2006Icar..185..113M,2008GeoRL..35.7202W}. However, so far data assimilation has not been applied to the thermal infrared data sets gathered from small solar system bodies.

 \section{Methods}
 The method described in this section is one of the standard approaches for nonlinear high dimensional state estimation. Here, "state" denotes the variables that describe the time-dependent condition of the system, i.e. the surface and sub-surface temperature, as opposed to "parameters" that govern the state, i.e. the thermophysical properties of the surface material.
 
  The DA technique is designed to infer states and parameters of a dynamical system of interest on the basis of two sources of information: a model (typically given by an evolution equation of a state of interest dependent on partially unknown parameters) and partial and noisy observations of the system. At first we will discuss the considered model and the associated observations followed by an introduction of the Ensemble Square Root filter \cite{jdw:tippett03,jdw:reichcotter15}.
 
\subsection{Model}

The thermophysical model used in this study is similar to the one used in \cite{Pelivan2017,hamm2018,grott2019} and assumes the surface to be a semi-infinite and homogeneous half-space. The 1D-heat conduction equation is solved
  \begin{equation}
    \label{eq:heat_conduction}
 	{
    \frac{\partial T(x,t)}{\partial t} =
    \frac{\pi}{\Omega}\frac{\partial^2 T(x,t)}{{\partial x^2}}
    }
  \end{equation}
 where $\Omega$ is the rotation period, $T(x,t)$ is the time- and depth-dependent temperature with $x$ being the depth variable in the direction of the local surface normal and $x = 0$ at the surface. The depth is normalised to the diurnal skin depth $d$ which is defined as:
  \begin{equation}
\label{eq:skindepth}
{
	d = \sqrt{\frac{k}{c_p \rho}\frac{\Omega}{\pi}}
	}
	\end{equation}
 where $\rho$ is the density, $c_{p}$ the specific heat capacity and $k$ is the thermal conductivity. This normalisation requires to assume $k$, $c_p$, $\rho$ to be constants. At the lower boundary the flux is set to zero. The upper boundary condition is given by the energy balance at the surface:
  	  \begin{equation}
	\label{eq:upper_bc_gamma}
	{
	(1-A)I(t) = \epsilon\sigma_B T^4(x=0,t) + \Gamma \sqrt{\frac{\pi}{\Omega}} \frac{dT}{dx}(x=0,t) + Q_{th}(t)
	}
	\end{equation}
where $A$ is the surface bond albedo, $I(t)$ is the solar illumination, $\epsilon$ is the thermal emissivity and $\sigma_B$ is the Stefan-Boltzman constant. Further $Q_{th}(t)$ denotes the thermal radiation received from the surrounding terrain. The thermal inertia is defined as $\Gamma = \sqrt{\rho c_p \kappa}$  in units of $\mathrm{J\,m^{-2}\,K^{-1}\,s^{-1/2}}$. This parameter is commonly used to describe the amplitude of the diurnal surface temperature variation and its phase shift with respect to maximum insolation. The higher a surface's thermal inertia the later it will reach its maximum temperature in the afternoon, and the smaller is the difference between day and night temperatures. 
This thermal model calculates the temperatures on a spatial grid. We chose this grid to consist of 41 points spread over eight diurnal skin depth with increasing distance, as described in \cite{hamm2018,grott2019}. 

The aim now is to determine unknown parameters of interest of Eq. (\ref{eq:upper_bc_gamma}), e.g. the thermal inertia $\Gamma$, emissivity $\varepsilon$, etc., which result in a specific temperature profile. 
In this paper we use a sequential data assimilation algorithm to simultaneously estimate the temperatures on the 41 grid points, the state, as well as the model parameters. This is achieved by defining an "augmented" state vector, which consists of the temperatures and model parameters and will be denoted $\mathbf{z}(t)\in\mathbb{R}^{N_z}$ in the following.

\begin{align}
    \label{eq:heat_conduction_augmented}
 	\mathbf{z}(t) = (T(0,t), T(x_1,t),...,T(x_{41},t),\Gamma,...)
\end{align}
where the temperatures evolve according to the thermal model Eq. (\ref{eq:heat_conduction}). Note that besides $\Gamma$, any parameter of the thermal model can be included in $\mathbf{z}$.

While the model parameters are time independent, the data assimilation requires some sort of evolution in time for a sequential improvement of the parameter's estimate. Here, a Brownian motion is chosen to ensure that the parameter space is traversed sufficiently to converge to the true parameter value. Here, the forward model of the thermal inertia is defined as
 \begin{align}
    \label{eq:parameter_evo}
    \frac{d \Gamma}{{d t}}&=\frac{dW_t}{dt}
 \end{align}
where $W_t$ is a Wiener process, i.e. the mathematical description of the Brownian motion. This process is realised by adding, in each forecast step, a random number to the previous estimate:
  \begin{align}
    \label{eq:parameter_evo2}
\Gamma(\tau_n)=\Gamma(\tau_{n-1}))+\zeta_{\Gamma}(\tau_{n-1}), \quad \textrm{with} \quad \zeta_{\Gamma} \sim \mathcal{N}(0,\alpha_{\Gamma}(\tau_{n-1})^2)
\end{align}
where $\tau_n$ is the time at which the parameter is updated, $\tau_{n-1}$ is the time of the previous update, and  $\zeta_{\Gamma}(\tau_{n})$ is a random number drawn from a normal distribution $\mathcal{N}$ centred on zero and with a standard deviation of $\alpha_{\Gamma}(\tau_n)$. The update time $\tau$ in this study is equivalent to the time of the observations.
This formalism can be applied to other parameters of the thermal model, each with their own choice of $\alpha$ as provided in Table \ref{tab:alpha}.

\subsection{Data}
In order to employ DA techniques, observations that can be linked to the state of interest are required. The relationship between observations $y_{\text{obs}}$ and augmented state $\mathbf{z}(\tau)$ can be written as
 \begin{equation}\label{eq:observations}
     y_{\text{obs}}(\tau_n)=\mathbf{H}\,\mathbf{z}(\tau_n)+\nu(\tau_n)
 \end{equation}
 where $\nu(t)\in\mathbb{R}^{N_y}$ is the observational noise which is assumed to be Gaussian distributed with zero mean and covariance matrix $\mathbf{R}\in\mathbb{R}^{N_y\times N_y}$ and $\mathbf{H}\in\mathbb{R}^{N_y\times N_z}$ is the observation operator. Further note that the number of observed components $N_y$ is often significantly smaller than the dimension of the augmented state space $N_z$.
 In this study only one component of $\mathbf{z}(\tau)$ is observable. In the first part of the study, this is the surface temperature $T(0,\tau)$. In the second part, it is the radiance emitted by the surface and observed by the MASCOT radiometer. This means that $N_y = 1$, $\mathbf{R}$ is a scalar corresponding to the measurement uncertainty, and  $\mathbf{H}$ is given by
 \begin{equation}
     \mathbf{H}(1) = 1 \quad \textrm{and} \quad \mathbf{H}(i) = 0 \quad \textrm{for} \quad i \neq 1.
 \end{equation}
 
\subsection{Sequential Data Assimilation and the Ensemble Square Root Filter}

In the following, we will briefly introduce basic concepts of sequential data assimilation and the classic Kalman filter as an example for sequential data assimilation.  Then, we will introduce the Ensemble Square Root Filter (ESRF) which is a prominent member of the family of Ensemble Kalman Filters (EnKFs)
\cite{evensen2006}.

\subsubsection{Classic Kalman Filter}

In sequential data assimilation, the probability distribution of a system's state is estimated by repeatedly applying two steps called forecast and analysis. Based on an initial estimate of the state, the state in the first time step is predicted by applying a model, e.g., the thermophysical model described above. This prediction is corrected by incorporating an observation. The corrected prediction is referred to as analysis and utilised as the input of the model to predict the state in the next time step. One iterates over this procedure for all subsequent time steps. The forward model can be expressed by:
 \begin{equation}\label{evolutionmodel}
  \mathbf{z}(\tau_n)=\Psi\Big(\mathbf{z}(\tau_{n-1})\Big)
 \end{equation}
 where $\Psi$ is the operator evolving the augmented state from time $\tau_{i-1}$ to $\tau_i$. For the temperature evolution, $\Psi$ is the solution to the PDE given in Eq. (\ref{eq:heat_conduction}) evolving the temperatures $T(x,\tau_{n-1})$ to the temperature forecast $T(x,\tau_n)$. For the  evolution of the model parameters, $\Psi$ is described in Eq. (\ref{eq:parameter_evo2}). Note that it is possible to add some noise in Eq. (\ref{evolutionmodel}) to express uncertainties stemming from model or numerical errors.

The concept of sequential data assimilation can be best illustrated by its standard version, the classic Kalman filter (KF) \cite{jdw:Kalman1960}. It is valid for a linear dependence of the observation on the state $\mathbf{z}$ and a linear forward model, i.e. a linear $\Psi$:
 \begin{equation}\label{lin_evolutionmodel}
  \mathbf{z}(\tau_n)=\mathbf{z}(\tau_{n-1})+\delta \tau (\mathbf{A}\,\mathbf{z}(\tau_{n-1})+\mathbf{b})
 \end{equation}
 where $\delta \tau$ is the time step, $\mathbf{A}$ and $\mathbf{b}$ are model parameters.
A common example is the estimation of the position of some vehicle based on measurements of velocity and position at given points in time \cite{hu2003adaptive}. 
The KF has the underlying assumption that the posterior distribution \[
 p(\mathbf{z}(\tau_n)|y_{\text{obs}}(\tau_1:\tau_n))
 \]
 that describes the probability of the augmented state $\mathbf{z}(\tau_n)$ given all the data $y_{\text{obs}}(\tau_1:\tau_n)$ from time $\tau_1$ up to time $\tau_n$ is a Gaussian $N(\mathbf{m}^a(\tau_n),\mathbf{P}^a(\tau_n))$ where $\mathbf{m}^a\in\mathbb{R}^{N_z}$ is the first and $\mathbf{P}^a\in\mathbb{R}^{N_z\times N_z}$ the second moment of the associated normal distribution. 
 
\begin{figure}
    	\centering
    	\includegraphics[width=\linewidth]{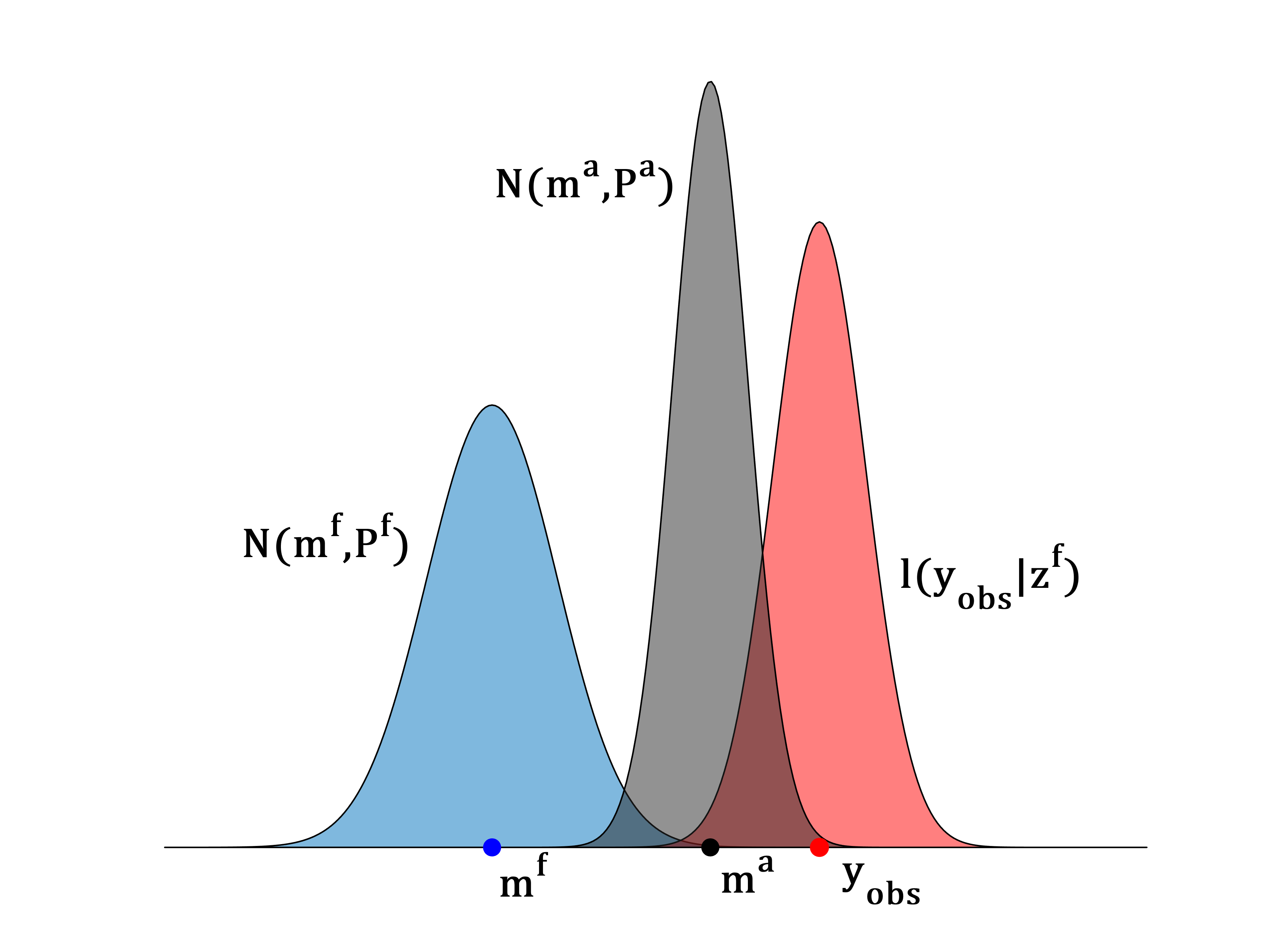}
    	\includegraphics[width=\linewidth]{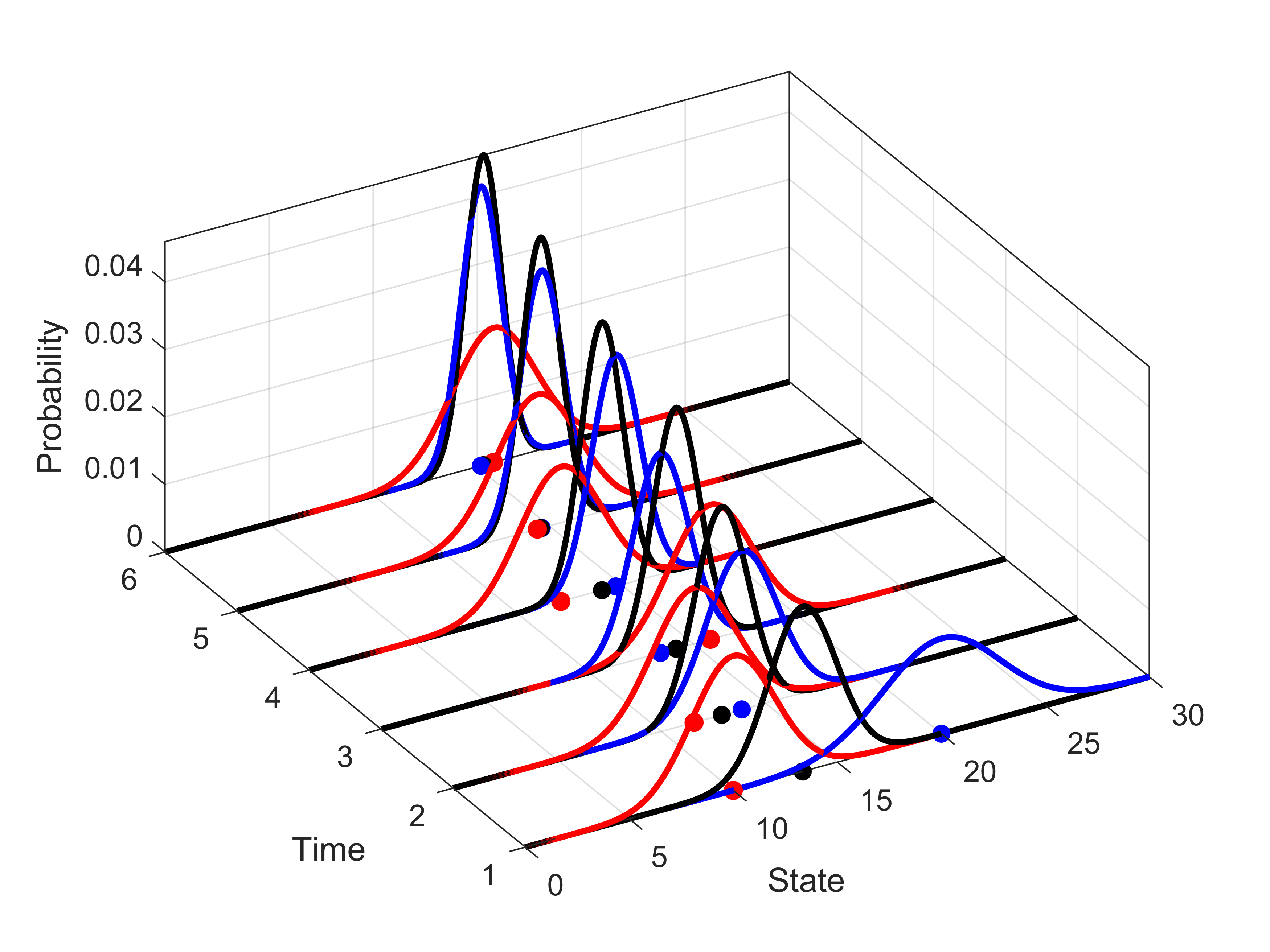}
    	
    	\caption{Top: The graph visualises the classic Kalman update. The blue (left) normal distribution is the \textit{prior} with mean $m^f$, while the red (right) represents the \textit{likelihood} function of the observation (red dot $y_{obs}$) and the black (centre) distribution describes the so called \textit{posterior} distribution with mean $m^a$. Bottom: Illustration of a sequence of Kalman updates. The analysis of one time step is used to forecast the state in the next time step. Mean of forecast and analysis are shown as blue and black dots respectively, the observation is shown as a red dot.
    	}
    	\label{fig:KL}
    \end{figure}

  Bayes Theorem connects this posterior distribution with the prior distribution which describes the probability of the augmented state $\mathbf{z}(\tau_n)$ given all the data $y_{\text{obs}}(\tau_1:\tau_{n})$ from time $\tau_1$ up to time $\tau_{n}$. The prior distribution is assumed to be Gaussian $N(\mathbf{m}^f(\tau_{n}),\mathbf{P}^f(\tau_{n}))$ as well. This is the case when the model operator $\Psi$ and the observation operator $\mathbf{H}$ are linear, and the initial value, the model and observational noise are independent identical Gaussian distributed. 
  The superscript $a$ and $f$ are abbreviations of \textit{analysis} and \textit{forecast} used to distinguish between posterior and the prior distribution respectively. 
 This notation is in accordance with the classical DA notation in the main application areas such as numerical weather prediction and oil recovery.
  The link between prior and posterior is achieved via the likelihood $l(y_{\text{obs}}(\tau_{n})|\mathbf{z}^f(\tau_n))$ which describes the probability of the observations conditioned on the current state estimate, i.e.,
 \begin{equation}
 \begin{aligned}
 \label{eq:Bayes}
 N(&\mathbf{m}^a(\tau_n),\mathbf{P}^a(\tau_n))\\ &\propto l(y_{\text{obs}}(\tau_{n})|\mathbf{z}^f(\tau_n))N(\mathbf{m}^f(\tau_n),\mathbf{P}^f(\tau_n))
 \end{aligned}
 \end{equation}
 The upper panel in Fig. \ref{fig:KL} shows the three probability distributions, prior, posterior and likelihood.

 The posterior distribution in Eq. (\ref{eq:Bayes}) is given by
 \begin{equation}
 \begin{aligned}
     \label{eq:posterior_explicit}
    N(&\mathbf{m}^a(\tau_n),\mathbf{P}^a(\tau_n) \\& \propto 
    \exp{} \Big( -\frac{1}{2}\Big( (y_{\text{obs}}(\tau_n)-\mathbf{H}\mathbf{z}^f(\tau_n))^{\top}\mathbf{R}^{-1}(y_{\text{obs}}(\tau_n)-\mathbf{H}\mathbf{z}^f(\tau_n))+ \\
    & (\mathbf{z}^f(\tau_n)-\mathbf{m}^f(\tau_n))^{\top}\mathbf{P}^{f-1}(\tau_{n})(\mathbf{z}^f(\tau_n)-\mathbf{m}^f(\tau_n)) \Big) 
    \Big)
 \end{aligned}
 \end{equation}

 The first part of the right hand side of the expression is the likelihood, $l(y_{\text{obs}}(\tau_{n})|\mathbf{m}^f(\tau_n))$, which is maximal when the forecast observation $\mathbf{H}\mathbf{z}^f$ is close to the observation $y_{\text{obs}}$. The second part is the prior distribution which is maximal at its mean $\mathbf{m}^f$.
 One can rearrange the exponent to show that the posterior is a Gaussian with mean and covariance given by
 \begin{equation}
 \begin{aligned}
     \label{eq:Kalman_posterior}
     \mathbf{m}^a(\tau_{n}) & = \mathbf{m}^f(\tau_{n}) -  \mathbf{K} (\mathbf{H}\mathbf{m}^f(\tau_{n})-y_{\text{obs}}(\tau_{n}))\\
     \mathbf{P}^a(\tau_{n}) & = \mathbf{P}^f(\tau_{n}) -  \mathbf{K} \mathbf{H} \mathbf{P}^f(\tau_{n})
 \end{aligned}     
 \end{equation}
 where $\mathbf{K}$ is the Kalman gain defined as 
 \begin{equation}
     \label{eq:K}
      \mathbf{K}(\tau_{n}) = \frac{\mathbf{P}^f(\tau_{n})\mathbf{H^{\top}}}{\mathbf{R}+\mathbf{H}\mathbf{P}^f(\tau_{n})\mathbf{H^{\top}}}
 \end{equation}
 
 The Kalman gain weights how much the analysis, i.e., the posterior distribution, is governed by the forecast produced by the model or the observation. Details of this derivation are given in chapter 6 of \cite{jdw:reichcotter15}. The smaller the observation error $\mathbf{R}$ the larger $\mathbf{K}$ becomes and the more is the observation weighted into the calculation of the analysis.
 Contrarily, if the observation error is very large, i.e.,

  \[  \mathbf{K} \approx 0 \quad (\text{ for  }\mathbf{R}>>\mathbf{P}^f)
   \]
 
 and consequently $\mathbf{m}^a =\mathbf{m}^f$ and $\mathbf{P}^a = \mathbf{P}^f$.
 One can show that the updated mean $\mathbf{m}^a$ maximises Eq. (\ref{eq:posterior_explicit}). One can furthermore show that $\mathbf{m}^a$ is the "best linear unbiased estimator" of the state for a linear system, i.e. it minimises the expectation value $\mathbb{E}\left[ \left\| \mathbf{m}^a - \mathbf{z}_{\text{true}} \right\|^2 \right]$.
 The mean and covariance of the updated posterior are used to forecast the state in the next time step by applying the linear forward model (Eq. \ref{lin_evolutionmodel}).
 \begin{equation}
 \begin{aligned}
     \label{eq:lin_mean_p_forward}
     \mathbf{m}^f(\tau_{n+1}) & = (\mathbf{I}+\delta \tau \mathbf{A})\mathbf{m}^a(\tau_n) + \delta \tau \mathbf{b} \\
     \mathbf{P}^f(\tau_{n+1}) & = (\mathbf{I}+\delta \tau \mathbf{A})\mathbf{P}^a(\tau_n) (\mathbf{I}+\delta \tau \mathbf{A})^{\top}
 \end{aligned}     
 \end{equation}
 
Iterations over Eq. (\ref{eq:Kalman_posterior}) and (\ref{eq:lin_mean_p_forward}) provide a sequence of best linear unbiased estimators for a series of observations for systems with linear dynamics, which is illustrated in the lower panel of Fig. \ref{fig:KL}.

\subsubsection{Ensemble Square Root Filter}
The classic KF can be extended to a nonlinear model setting via an ensemble approach where an ensemble of $M$ augmented state vectors $\mathbf{z}^f_i(\tau_n)$ and $\mathbf{z}^a_i(\tau_n)$ with $i\in\{1,\dots,M\}$ are generated in each time step $\tau_n$ to approximate the Gaussian prior and posterior distribution via the empirical posterior mean
  \begin{equation}
\hat{\mathbf{m}}^a(\tau_n)=\frac{1}{M}\sum^M_{i=1} \mathbf{z}^{a}_{i}(\tau_n)
 \end{equation}
and covariance
 \begin{equation}\label{eq:empiricalP}
\hat{\mathbf{P}}^{a}(\tau_n)=\frac{1}{M-1}\sum^M_{i=1} (\mathbf{z}^{a}_{i}(\tau_n)-\hat{\mathbf{m}}^{a}(\tau_n))(\mathbf{z}^{a}_{i}(\tau_n)-\hat{\mathbf{m}}^{a}(\tau_n))^{\top}
 \end{equation}
 for each $\tau_n$ and analogously for the empirical prior distribution. 
 This means that the Ensemble Kalman filters are Monte-Carlo approximations of the classic KF. As in the classic KF the forecast prior is updated incorporating the observation. The mean and covariance of the analysis have to fulfil the Kalman Update in Eq. (\ref{eq:Kalman_posterior}). However, while in the KF it was sufficient to update mean and covariance, in the ensemble Kalman filter each "ensemble member", i.e. $\mathbf{z}^f_i$ has to be updated individually.
 
 The iterative update procedure of the samples is described in Algorithm \ref{alg:ESRF}. The corresponding code will be made available upon request. At first the initial ensemble of $M$ augmented state vectors $\mathbf{z}_i(0)$ with $i\in\{1,\dots,M\}$ are generated by sampling from Gaussian distributions $N(\mathbf{m}(0),\mathbf{P}(0))$ which are then, individually, sequentially updated by iterating over the forecast and analysis step.

\begin{algorithm}
\caption{Ensemble Square Root Filter}\label{alg:ESRF}
\begin{algorithmic}
\State Set variables $\mathbf{m}(0)$, $\mathbf{P}(0)$ and $M$
\State Initialise ensemble of augmented states $\mathbf{z}_i(0)\sim N(\mathbf{m}(0),\mathbf{P}(0))$ with $i \in\{1,\dots,M\}$ by means of $N(\mathbf{m}(0),\mathbf{P}(0))$
\State 
	\For {$n=1:N$}
			\State  \textbf{Forecast:}
			     \begin{equation}
\mathbf{z}^f_i(\tau_n) =\Psi\Big(\mathbf{z}^a_i(\tau_{n-1})\Big)\quad \forall i\in\{1\dots,M\}
 \end{equation}

				\State \textbf{Analysis step:} 
			        \begin{equation}
        \mathbf{z}_i^a(\tau_n)  = \sum^M_{j=1} D_{ji} \mathbf{z}^f_j(\tau_n) \quad 
        \end{equation}
        $\text{with update }D_{ji} \text{ given in Eq. (\ref{eq:dij})}$
	\EndFor
\end{algorithmic}
\end{algorithm}

   The update matrix $\mathbf{D}\in\mathbb{R}^{M\times M}$, which performs the update for each ensemble member, is constraint by the condition that after the update the mean and covariance of the ensemble members fulfil the Kalman update and the calculation of $\mathbf{D}$ depends on which EnKF variant is used. The different versions can be divided into a stochastic branch and a deterministic one. For this study we choose the deterministic branch, the Ensemble Square Root Filter approach (ESRF).
   The numerical success of the family of EnKFs has been documented for various applications \cite{evensen2006} and there are rigorous accuracy and stability analyses available for the considered ESRF  \cite{deWiljesTong2019,deWiljesStannatReich2019}. Further the proposed deterministic branch of this family is preferable \cite{jdw:tippett03,Nergeretal2012} over the stochastic alternative (also know as perturbed EnKF) which is also very popular in the literature \cite{Evensen2003}.
   The entries 
     \begin{equation}
    \label{eq:dij}
        D_{ji} = w_j -\frac{1}{M} + S_{ji}
    \end{equation}
    of $\mathbf{D}$ depend on the the components $w_j$ of the weight vector
    \begin{equation}
        \mathbf{w} = \frac{1}{M}\mathbf{1}- \frac{1}{M-1} \mathbf{S}^2 (\mathbf{E}^f)^{\top} \mathbf{H}^{\top}\mathbf{R}^{-1}(\mathbf{H}\hat{\mathbf{m}}^f-y_{\text{obs}})\in\mathbb{R}^{M}
    \end{equation}
    where $\mathbf{1}$ is a column vector filled with ones and in $\mathbb{R}^{M}$. $\mathbf{E}^f$ is a matrix in $\mathbb{R}^{N_z\times M}$ that provides the distance between each ensemble member to the mean of the ensembles:
    \begin{equation}
        \mathbf{E}^f = [(\mathbf{z}_1^f - \hat{\mathbf{m}}^f) \quad ... \quad (\mathbf{z}_M^f - \hat{\mathbf{m}}^f)]
    \end{equation}
    The matrix $S$ and its entries $S_{ij}$ that enter Eq. (\ref{eq:dij}) are defined by:
    \begin{equation}
    \label{eq:s}
        \mathbf{S} = \Big(\mathbf{I} + \frac{1}{M-1}(\mathbf{HE}^f)^{\top}\mathbf{R}^{-1}\mathbf{HE}^f\Big)^{-1/2}
    \end{equation}

where the matrix square root is defined as $\mathbf{B}^{1/2}\mathbf{B}^{1/2}=\mathbf{B}$ for a matrix $\mathbf{B}$. The name "Ensemble Square Root Filter" refers to this matrix square root computation. 

Note, that the update occurs at the times $\tau$, i.e., the observation times. For the sake of readability the time dependency is not explicitly written in Eq. (\ref{eq:dij})-(\ref{eq:s}).
The update matrix $\mathbf{D}$ is constructed so that empirical mean and covariance are equal to the true mean and covariance of the posterior, $\mathbf{m}^a=\hat{\mathbf{m}^a}$ and $\mathbf{P}^a=\hat{\mathbf{P}^a}$, for a linear $\Psi$. In other words the algorithm is designed to produce the same mean and covariance as the classic KF for a linear setting even for finite number of ensemble members $M$, whereas other EnKF versions only produce the KF mean and covariance in the ensemble limit $M\rightarrow\infty$, e.g., the perturbed EnKF \cite{jdw:EvensenLeeuwen2000}. Further note that the update of each ensemble member depends on all other ensemble members, coupled through the empirical covariance matrix $\hat{\mathbf{P}^a}$ given in Eq. (\ref{eq:empiricalP}). For a more detailed derivation of the ESRF and its properties see chapter 7 in \cite{jdw:reichcotter15}.

This form of update does not require the model to be linear which is one of the key benefits of the ESRF compared to the classic KF. Furthermore, despite the underlying Gaussianity assumption, the nonlinear evolution of the particles allows to capture the more complex behaviour of the system and thus leads to more realistic estimates.

\section{Numerical simulation}

The ESRF is tested for two cases. The first case is a proof-of-concept where the thermal inertia is derived in a controlled and simplified set-up with an artificial dataset based on a reference solution of the thermophysical model. In the second case it is employed to revisit the analysis of the radiometric data set retrieved by the MASCOT lander from the surface of Near-Earth asteroid (162173) Ryugu \cite{grott2019}.

 \subsection{Estimation of Thermal Inertia in a Simplified Model}
 The aim of this numerical example is to show how the technique performs in a controlled setting. This is achieved by generating an artificial reference temperature profile computed by means of a set of fixed reference parameters. In order to validate the performance of the proposed technique the estimates obtained via the ESRF are compared to the reference temperature variation and reference thermal inertia.
 
 \subsubsection{Reference Solution}
The reference temperature is simulated using the model given in Eq.  (\ref{eq:heat_conduction}) above with thermal inertia $\Gamma^{\text{ref}}=300$ $\mathrm{J\,m^{-2}\,K^{-1}\,s^{-1/2}}$, an albedo of 0.015, emissivity of 1, and $Q_{th}$ of 0. The illumination is calculated by the simple assumption of a spherical asteroid, with equal length of day and night:

	  \begin{equation}
	\label{eq:illu_case1}
	{
	I(t) = I_{max} \cos{(\frac{2\pi}{\Omega} \, t)} 
	}
	\end{equation}
where $I(t) = 0$ if $\cos{\frac{2\pi}{\Omega} t} < 0$ and $I_{max} = 800 \, \mathrm{W/m^{2}}$ similar to the illumination conditions on Ryugu. The rotation period $\Omega$ can be chosen arbitrarily and was set to the rotation period of Ryugu of $7.63262$ h \cite{watanabe2019}.

\subsubsection{Initial ensemble}

For each ESRF simulation an initial ensemble of $M=50$ members is generated and the thermal inertia values of these ensemble members are then drawn from a Gaussian distribution, $\Gamma_i = \Gamma^{start}+\zeta_i$, with $\zeta_i \sim \mathcal{N}(0,\alpha^2)$ and $\alpha = 20$  $\mathrm{J\,m^{-2}\,K^{-1}\,s^{-1/2}}$. We repeat this procedure for $20$ ESRF simulations, sampling $\Gamma^{start} \sim \mathcal{N}(250,100^2)$ rather than running a single simulation of $1000$ Members with a standard deviation of 100 $\mathrm{J\,m^{-2}\,K^{-1}\,s^{-1/2}}$. We found that by doing so we gain a more homogeneous sampling of the initial distribution in parameter space. Furthermore, we save computation time as each of the $20$ ESRF simulations converges quicker than a single simulation with a larger ensemble and the individual runs can be evaluated in parallel.

In order to save more computation time, a number of temperature profiles are pre-calculated assuming the parameters given above and varying the thermal inertia between $100$ and $500$ $\mathrm{J\,m^{-2}\,K^{-1}\,s^{-1/2}}$ in steps of 50 $\mathrm{J\,m^{-2}\,K^{-1}\,s^{-1/2}}$. The ensemble member's initial temperature profiles are than initialised by interpolating from the pre-calculated temperature profiles. These provided a more realistic initial guess for the temperature solution, accelerating the convergence of the PDE-solver. 

For each ensemble member the temperature profile is sampled from a Gaussian distribution centred on the interpolated temperature profile with a standard deviation of 1 K. This method ensures that the ensemble is spread sufficiently to evolve through the parameter space while at the same time keeping the temperature profiles close enough to a physical solution to ensure convergence of the differential equation solver. 

\subsubsection{DA settings}

The partial differential equation is solved using the MATLAB\textsuperscript{\textregistered} "pdepe"-solver for a total of $30,000$ time steps per simulated rotation, i.e. diurnal cycle. This corresponds to a time step $\Delta t = 0.91$ s. 

For the Kalman update, $15$ observation points are placed equidistantly in time from noon ($\tau = 0$) to noon. The thermal model is run between these observations for $2000$ $\Delta t$ to forecast the temperature profile at the next observation, using the thermal inertia from the last Kalman update.
The observation error is set to $1$ K, which corresponds to setting the associated covariance matrix to $\mathbf{R}=1$. 

The augmented state vector is then given by 
\begin{equation}
\label{eq:augstate1}
\mathbf{z}(\tau) = (T(0,\tau), T(x_1,\tau),...,T(x_{41},\tau),\Gamma(\tau)) .   
\end{equation}
The thermal inertia evolves as described in Eq. (\ref{eq:parameter_evo}) and (\ref{eq:parameter_evo2}), where the parameter $\alpha(\tau)$ is reduced from one simulated rotation to the next. The width is varied from $\alpha = 10$ $\mathrm{J\,m^{-2}\,K^{-1}\,s^{-1/2}}$ in the first rotation to $\alpha = 5$ $\mathrm{J\,m^{-2}\,K^{-1}\,s^{-1/2}}$ in the second, $\alpha = 1$ $\mathrm{J\,m^{-2}\,K^{-1}\,s^{-1/2}}$ in the third, $\alpha = 0.5$ $\mathrm{J\,m^{-2}\,K^{-1}\,s^{-1/2}}$ in the fourth, and $\alpha = 0.2$ $\mathrm{J\,m^{-2}\,K^{-1}\,s^{-1/2}}$ for the remaining 16 rotations. This gradual decrease of $\alpha$ is in line with classic simulated annealing schedules often employed in the context of Monte Carlo methods. The key idea however is very intuitive, i.e., big $\alpha$ help to traverse the parameter space more quickly while they also prevent convergence of the ensemble members. Thus lower $\alpha$ values are chosen as the estimation procedure progresses in order to allow the posterior distribution to converge.

\begin{figure}
    	\centering
    	\includegraphics[width=\linewidth]{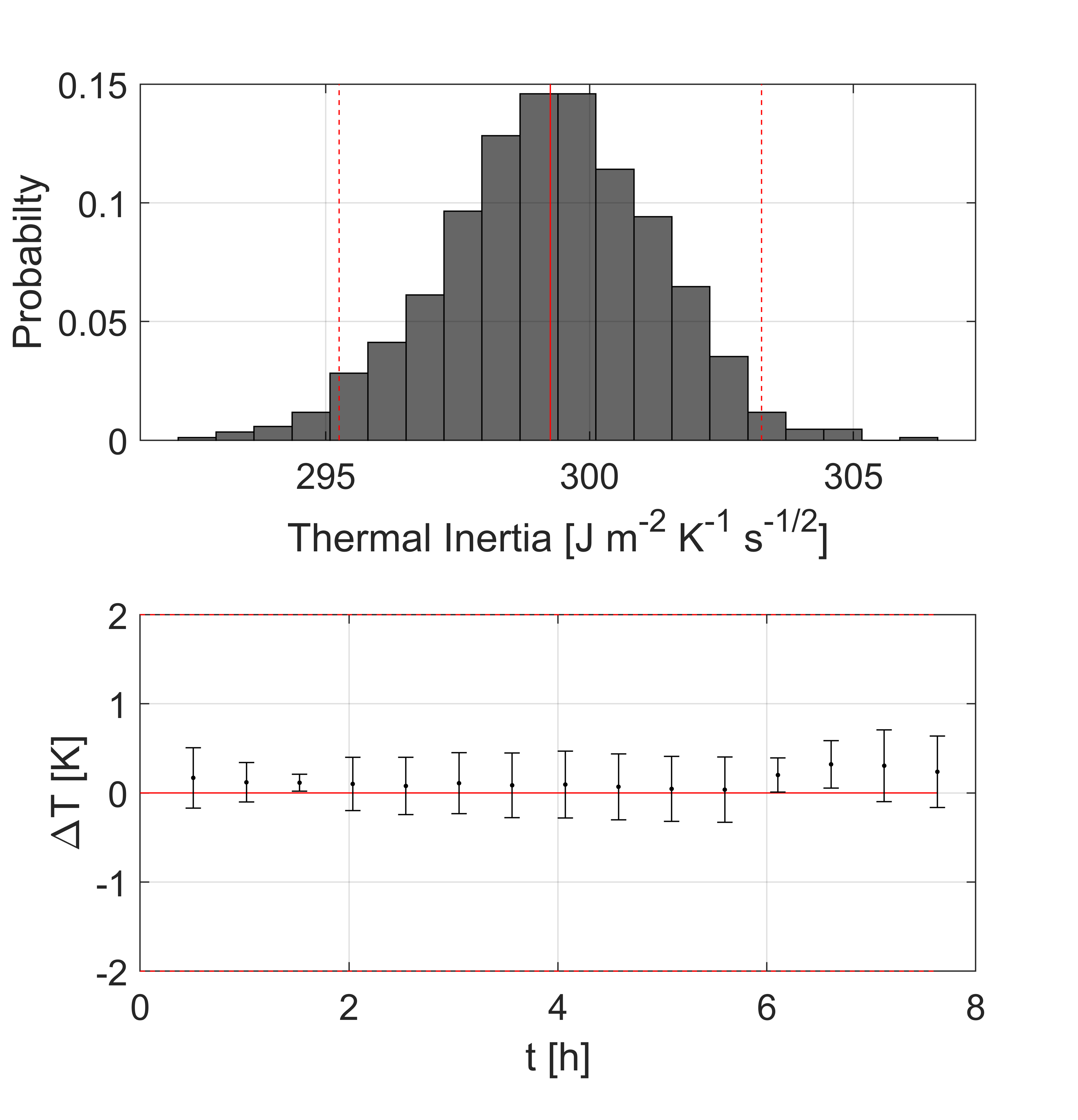}
    	\caption{Top: Histogram of the thermal inertia of all the ensemble members from the last time step after twenty simulated rotations. Y-axis shows the probability, i.e. the number of elements in a bin divided by the total number of estimates ($1000$). The solid red line indicates the mean, dotted red lines indicate $2\sigma$, with $\sigma$ denoting the standard deviation. Bottom: Black symbols indicate the mean of the deviation between the reference solution and the solution of the ensemble members in the last simulated diurnal cycle. The error bar indicates $2\sigma$, with $\sigma$ being the standard deviation over all ensemble members and simulations. The temperature estimates lie well within the postulated 2 K observation error for the $2\sigma$ uncertainty.}
    	\label{fig:TIonly}
    \end{figure}

\subsubsection{Results}
     The $20$ ESRF simulations were run for $20$ asteroid rotations starting with a randomly chosen thermal inertia each. Fig. \ref{fig:TIonly} shows the histogram of the thermal inertia after 20 rotations. During the first few simulated cycles, the parameters spread out wide before converging. The last thermal inertia estimates of all ensemble members over all $20$ simulations were combined into the histogram, i.e. 1000 thermal inertia estimates make up the final result of $\Gamma = 299 \pm 4$ $\mathrm{J\,m^{-2}\,K^{-1}\,s^{-1/2}}$ with the uncertainty given by the $2\sigma$ bound, where $\sigma$ is the standard deviation over the thermal inertia set. The thermal inertia of the reference temperature was 300 $\mathrm{J\,m^{-2}\,K^{-1}\,s^{-1/2}}$ and could thus be successfully retrieved.
    
    Furthermore, the reference temperature could be retrieved well within the assumed 1 K uncertainty. The bottom panel of Fig. \ref{fig:TIonly} shows the temperature estimates at the 15 observation points, where at each point the mean and standard deviation was taken over the last diurnal cycle. The error bar indicates the $2\sigma$ uncertainty.
    
    This study demonstrates the working principle of the considered data assimilation algorithm for the retrieval of thermophysical parameters from temperature observations. In the next step the ESRF will be used to revisit the radiometric data obtained on the surface of asteroid (162173) Ryugu by the MARA instrument \cite{grott2017,grott2019}.

 \subsection{Thermal Inertia Estimation of Ryugu}
 
 The MARA instrument onboard the MASCOT lander observed the infrared flux emitted by the surface of a single, irregularly shaped boulder on Ryugu for a full diurnal cycle. The instrument consists of six infrared bolometers, that are placed behind different infrared filters. The 8 - 12 \textmu{}m (W10) filter was the instrument channel with the highest fidelity and was used for the initial analysis reported in \cite{grott2019}. 
 In that work, the nighttime data was fitted by minimising the $\chi^2$ value that measured the misfit between observed flux and the one predicted by a thermal model. 
 
 This thermal model included as free parameters the thermal inertia $\Gamma$, emissivity $\varepsilon$, the orientation of the surface observed by MARA in terms of azimuth $\theta$ and elevation $\phi$ of the surface normal, and the view factor to the surrounding terrain $f$ which parametrizes $Q_{th}$ as follows:

	\begin{equation}
	\label{eq:TRR}
	{
	Q_{th}(t) = f \sigma_B \varepsilon T_{obs}^4(t)
	}
	\end{equation}
  where the temperature of the surrounding is assumed to be equal to the brightness temperatures observed by MARA, $T_{obs}$, as described in \cite{grott2019}. 
It should be noted here, that the roughness correction applied to daytime observation in \cite{grott2019} did not influence the nighttime temperatures, and is therefore omitted in this study.
 
 The surface orientation had to be included as a free parameter as the observed spot on the irregular boulder showed a rugged texture with various parts of unknown orientation, and thus unknown illumination condition. The parameters $\theta$ and $\phi$ therefore represent an averaged surface orientation within the field of view of MARA.
 
 The illumination is determined by the scalar product of surface orientation $\mathbf{n}$ and (time-dependent) solar vector $\mathbf{s}$:
 \begin{equation}
     \label{eq:illu_ryugu}
     {
     I(t) = \frac{I_0}{r_h^2} \, \mathbf{n}(\theta,\phi) \cdot \mathbf{s}(t)
     }
 \end{equation}
  where $I_0$ is the solar constant and $r_h$ is the heliocentric distance.
 
 In \cite{grott2019}, the parameter space was sampled by a grid search, where the thermal inertia was varied in steps of 1 $\mathrm{J\,m^{-2}\,K^{-1}\,s^{-1/2}}$. However, due to the high computational cost of the thermal model, the other parameters were varied in significantly coarser steps, i.e. only three emissivity values were considered (0.9, 0.95, 1) along with only two values for $f$ ($0$ and $0.08$).
  To test the efficiency of our new approach this analysis was repeated using the ESRF.
 
 \begin{table}
    
    \begin{center}
    \begin{tabular}{lccccc}
    \hline
    $\alpha$ &$N_{rot}=1$ & $N_{rot}=2$  &  $N_{rot}=3$ & $N_{rot}=4$   & $N_{rot}>4$ \\
    \hline
    $\Gamma $ & 10 & 5 & 5 & 2 & 2\\
    $\varepsilon$ & 0.01 & 0.005 & 0.005 & 0.002 & 0.001\\
    $f$ & 0.001 & 0.0005 & 0.0005 & 0.0002 & 0.0001\\
    $\theta$ & 10 & 5 & 2 & 2 & 1\\
    $\phi$& 1 & 0.5 & 0.5 & 0.5 & 0.1\\
    \hline
    \end{tabular}
    \end{center}
   \caption{Overview of the parameter forecasts in the re-analysis of the MARA dataset. Listed are the standard deviations of the Gaussian distributions from which the parameter forecasts are sampled according to Eq. (\ref{eq:parameter_evo2})}
   \label{tab:alpha}
   \end{table}

 \subsubsection{Forecast settings}

    The forecasts of parameters and temperatures are again calculated using Eq. (\ref{eq:heat_conduction}) and a forward model of the free parameters as in Eq. (\ref{eq:parameter_evo}). The free parameters of the model were chosen analogous to the analysis of \cite{grott2019}: $\Gamma, \varepsilon,\phi,\theta,f$. Also, the same grid settings were applied, i.e. we calculate the temperature profiles for a $1$D grid with $41$ grid points spread over eight diurnal skin depths.
    The augmented state is then given  by:
    \begin{equation}
    \label{eq:augstate2} 
    \begin{split}
        \mathbf{z}(\tau) = & (F_{W10}(\tau),T(0,\tau), T(x_1,\tau), ... \\  & ...,T(x_N,\tau),\Gamma, \varepsilon,\phi,\theta,f).
    \end{split}
    \end{equation}

    Note that unlike in Eq. (\ref{eq:augstate1}) the surface temperature $T(0,t)$ is not directly observable but connected to the observed surface  radiance $F_{W10}$ via the instrument function.
        \begin{equation}
        \label{eq:radiance} 
        F_{W10}(\tau) = \varepsilon \int \mathrm{d}\lambda q(\lambda) B(T(0,\tau),\lambda)
    \end{equation}
  where $\lambda$ is wavelength, $B$ is the Planck function, and $q$ is the MARA filter throughput \cite{grott2017}. The radiance observed by MARA is calculated from the reported brightness temperatures \cite{grott2019}, i.e. $T$ in Eq. (\ref{eq:radiance}) is set to the brightness temperature and $\varepsilon = 1$. Note, that the MARA signal depends linearly on the net flux between MARA and the surface. The calculation of the brightness temperature from the signal incorporates the temperature of the MARA sensor, the instrument field of view, sensitivity, etc., which does not need to be repeated in this study.
  As in the simplified case, the forecast of the parameters is performed according to Eq. (\ref{eq:parameter_evo}) and (\ref{eq:parameter_evo2}). The value is again sampled from a Gaussian distribution where the standard deviation $\alpha(\tau)$ is stepwise reduced. An overview of $\alpha$ for the respective model parameters is provided in table \ref{tab:alpha}.
  
 	 \begin{figure*}
		\centering
				\includegraphics[width=\textwidth]{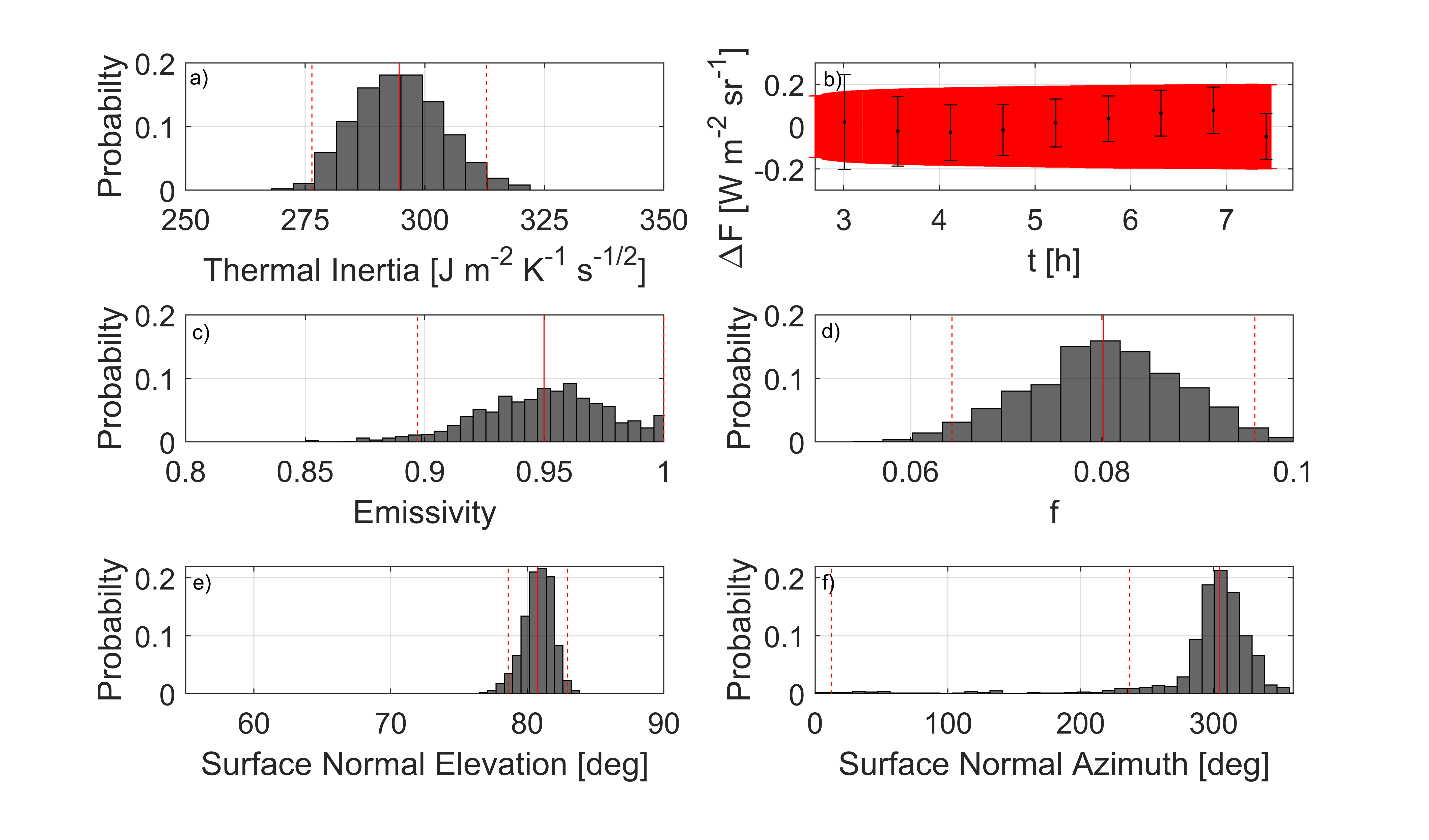}
		\caption{Parameter estimation from MARA dataset a) Histogram of thermal inertia estimates of ensemble members. The solid red line indicates the mean, the dotted red line indicates $2\sigma$, with $\sigma$ the standard deviation. b) Black: Deviation between the estimated radiance and the radiance emitted by the surface and observed by MARA as a function of time given in hours since the first MARA observation point (UTC 08:02:31), the errorbars indicate $2\sigma$ with $\sigma$ the standard deviation over all ensemble members and simulations. The estimated radiance is derived from the ensemble member's emissivity, surface temperature, and the instrument calibration. The red area indicates the $2\sigma$ uncertainty of the observed radiance based on the instrument calibration. The estimates of the other free parameters are shown in the following four histograms: c) emissivity d) integrated view factor  e) surface normal elevation f) surface normal azimuth. As in a), all  histograms are based on all ensemble members and simulations for the last Kalman analysis step.}
		\label{fig:MARA}
	\end{figure*}
 
\subsubsection{Data settings}

The Kalman updates are performed at nine points of the nighttime data equally distanced in time starting from 17:45 local time, which corresponds to the first data point considered in \cite{grott2019}. The distance between the points is similar to the one in the first part of this study, except for the last nighttime data point where the step to the first point in the next simulated night encompasses the entire asteroid day. The temperature is forecast by the thermal model, and converted into radiance received by the MARA W10 filter ($F_{W10}$), based on the instrument calibration \cite{grott2017,grott2019} and the ensemble member emissivity.

The illumination is calculated for each ensemble member based on the angle between the sun vector and the surface normal according to Eq. (\ref{eq:illu_ryugu}), while azimuth $\theta$ and elevation $\phi$ of the surface normal are updated from observation to observation. Likewise, surface emissivity, the view factor $f$, and the surface thermal inertia are updated.

\subsubsection{Initialisation}
     
 The ensemble states are initialised similar to the first case. In total $20$ ESRF simulations are performed and for each simulation a thermal inertia is randomly picked from $\Gamma^{start}_i \sim \mathcal{N}(300,100^2)$. In each simulation an ensemble with $50$ members is initialised, and for each ensemble member a thermal inertia is sampled from  $\Gamma_i \sim \mathcal{N}(\Gamma^{start},100^2)$. The thermal inertia is confined to an interval of 150 to 450 $\mathrm{J\,m^{-2}\,K^{-1}\,s^{-1/2}}$, a range that is larger than given by conservative estimates for Ryugu's thermal inertia \cite{wada2018}. For a sample of $\Gamma_i>450$, the thermal inertia is set to $450$, if $\Gamma_i<150$ it is set to 150.
 The emissivity of the ensemble members is sampled from a Gaussian distribution $\varepsilon \sim \mathcal{N}(1,0.02^2)$ and confined to the interval of $0$ and $1$. Thereby, emissivity values larger one are folded back into the interval, i.e. an $\varepsilon = 1.05$ is set to $0.95$ etc. 
 The view factor to the surrounding terrain $f$ is sampled from $f \sim \mathcal{N}(0.048,0.007^2)$ based on the topography of the landing site as described in the methods section of \cite{grott2019}.
 
 Azimuth and elevation of the surface normal in the best fitting case of the initial MARA data analysis were found to be $20^\circ$ and $80^\circ$  respectively, where an azimuth of $0^\circ$ is defined by the local east and an elevation $90^\circ$ corresponding to $314.207^\circ$ East and $34.599^\circ$ South in Ryugu's body fixed frame. For each ensemble member elevation and azimuth are sampled from $\theta \sim \mathcal{N}(20,360^2)$ where values are confined to 0 and $360^\circ$, e.g $\theta = 361^\circ = 1^\circ$ or $\theta = -1^\circ = 359^\circ$, and  $\phi\sim \mathcal{N}(80,10^2)$ where $\phi > 90^\circ$ are flipped back, e.g. a $\phi = 95^\circ$ is set to $\phi=85^\circ$ as an elevation larger $90^\circ$ is not defined.
 
 The temperatures of the ensemble members are initialised by interpolating the temperature profile from pre-calculated simulations. These pre-calculated simulations were performed for $\varepsilon=1$, $\theta=20$, $\phi=80$, and $f=0$, while the thermal inertia was varied between 150 and 450 $\mathrm{J\,m^{-2}\,K^{-1}\,s^{-1/2}}$ in steps of 50 $\mathrm{J\,m^{-2}\,K^{-1}\,s^{-1/2}}$. It should be noted here that the resulting initial temperature curves are not consistent with the initial parameter sets of the ensemble members. However, this is not problematic as the ensembles are given enough time to produce consistent solutions. Rather, this initial temperature profile serves as a better first guess for a solution of the 1D-heat conduction equation, e.g. compared to assuming a constant temperature as in \cite{grott2019}, and results in a quicker convergence of the temperature solution.
 
\subsubsection{Results}

    The data assimilation method allowed for a much finer variation of the free parameters, resulting in a more thorough estimate of the thermal inertia. 
    
    Fig. \ref{fig:MARA} shows parameter estimates of the ensembles at the last Kalman analysis step combining 20 simulations with randomly chosen starting thermal inertia. Mean and uncertainty of the estimates are displayed by solid and dashed red lines respectively with the uncertainty given as two standard deviations ($2\sigma$). The histograms displaying the posterior distributions of the various, simultaneously estimated model parameters show a major advantage of this method, which can account for non-Gaussian distributions.
    
    The thermal inertia was found to be $295 \pm 18$ $\mathrm{J\,m^{-2}\,K^{-1}\,s^{-1/2}}$. This result lies within the range of the former estimate but with lower uncertainty. The thermal inertia estimate is roughly Gaussian distributed, with a slight tilt towards lower values, accounting for the fact that the effect of thermal inertia on the temperature variation decreases with increasing thermal inertia. 
    
    The other parameter distributions contain important information about the observed boulder. The estimated emissivity is very high and estimated to be between $0.95\pm0.05$. This is in line with the extremely dark appearance of Ryugu and the fact that roughness, as observed on the boulder in front of MASCOT, tends to increase the emissivity even further. The estimates for $f$, $0.08\pm0.02$, show that this parameter might have been underestimated in \cite{grott2019}, i.e. that the view factor of the observed spot towards the surrounding terrain is up to $10\%$.
    
    The elevation of the average surface orientation within the MARA field of view is estimated to be between $81^\circ\pm2^\circ$, which is consistent with the camera images of the boulder surface in the field of view \cite{jaumann2019,scholten2019boulder}. The azimuth distribution shows that the most likely values lie within $304^\circ\pm68^\circ$, i.e. oriented towards south-east. This is also the direction of the MARA boresight, which is consistent with the fact that those surface parts oriented towards MARA will contribute most to the signal. Also, due to the roughness effect, such an orientation would result in a systematically lower noon temperature as reported by \cite{grott2019}. The former best-fit azimuth of $20^\circ$ is less likely as the very flat peak in the posterior distribution indicates. However, many of the fitting models reported in \cite{grott2019} showed an azimuth similar to the one retrieved in this work. Note that the dashed line in Fig. \ref{fig:MARA} f) at about $12^\circ$ represents the upper limit of the estimate as an azimuth angle of $372^\circ$ is equivalent to $12^\circ$.
    
    The figure also shows, that the estimated surface radiance matches the observed one very well. The major improvement of this analysis over the initial one is the full correlation among the parameter estimates and a smooth, simultaneous, and statistically thorough estimation rather than a rough parameter sweep. Despite the fact that $47$ parameters are estimated simultaneously, including the $40$ sub-surface temperatures, one simulation run requires only $30$ minutes on a Laptop with $4$ cores, drastically decreasing the computational resources needed.

        \begin{figure}
    	\centering
    	\includegraphics[width=\linewidth]{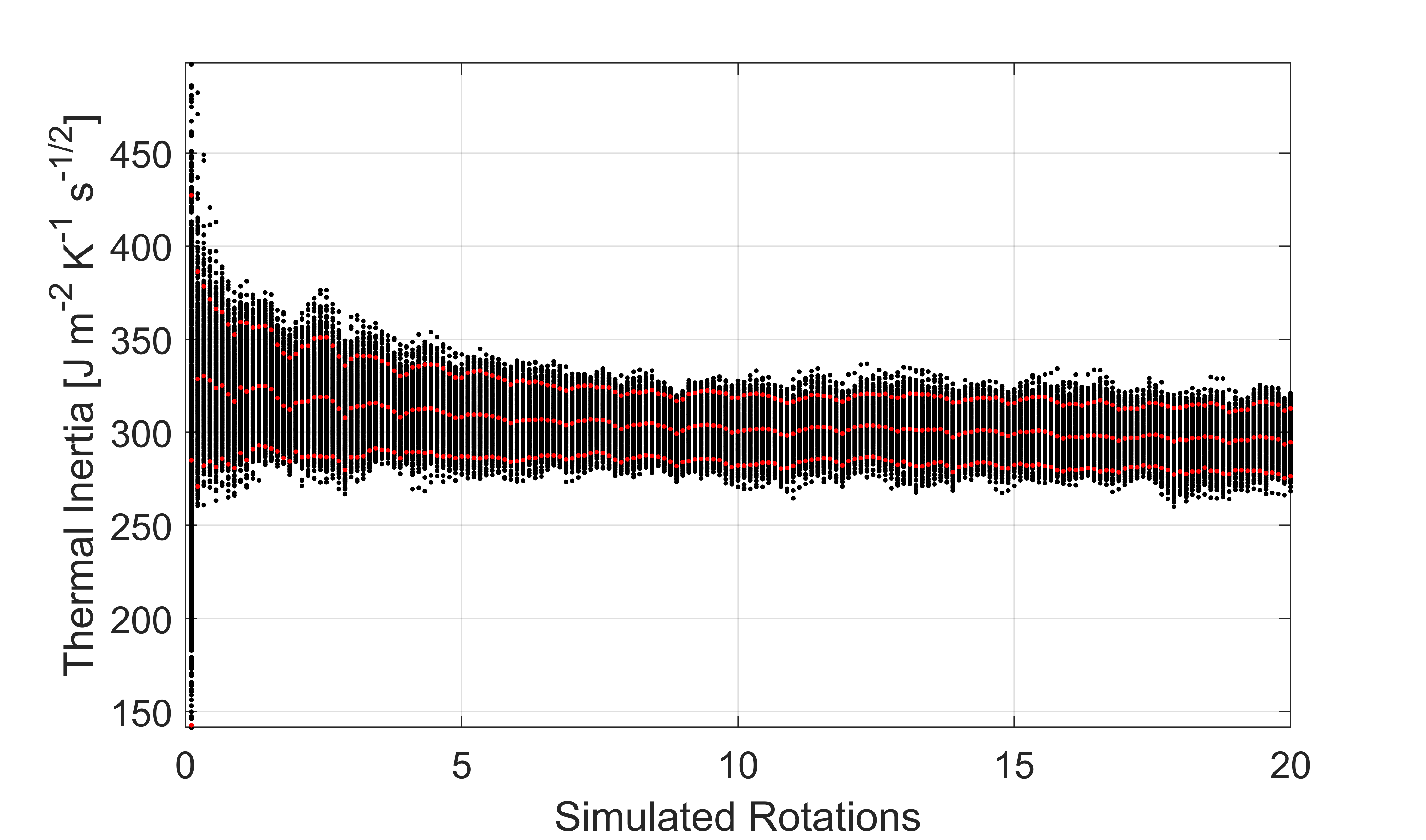}
    	\caption{Evolution of ensemble thermal inertia with time for all $20$ simulations ($1000$ ensemble members in total). Black dots indicate the thermal inertia of the ensemble members, red lines indicate the mean and standard deviation ($2\sigma$) in each time step.}
    	\label{fig:convergence}
    \end{figure}

    \subsection{Convergence of Ensemble Distribution}

    Since one of the major advantages of using an EnKF variant for the parameter estimation is the increased computational speed, it is important to investigate the convergence of the estimate. Fig. \ref{fig:convergence} shows the evolution of the ensemble thermal inertia.
    The mean and standard deviation are shown in red. The initial, wide-spread thermal inertia of the ensemble members quickly converges to the final ensemble spread. After 10 simulated rotations, the results change only slightly and after 15 simulated rotations the result is practically constant.
    
    To obtain a stable result, the number of simulations starting with different initial parameter combinations is more important than the length of each simulation. The combined results of 20 simulations, each with 50 ensemble members, converged quicker than the result of a single simulation (not shown in figure). Since the different simulation are independent of each other and can be run in parallel, this saves substantial computation time. 
    
    However, the most significant saving in computational cost is the efficient sampling of the parameters space. In a parameter sweep or also in other Monte-Carlo approaches, most of the tested parameter combinations have to be discarded, whereas the EnKF approach moves them through parameter space to a region of high probability. To obtain our results 20 simulations with 50 ensemble members were performed, i.e. 1000 model runs. A parameter sweep with comparable resolution, e.g. thermal inertia in steps of 5 from 250-400 $\mathrm{J\,m^{-2}\,K^{-1}\,s^{-1/2}}$, emissivity in steps of 0.01 from  0.8-1, $f$ in steps of 0.05 from 0 -  0.1, elevation in  steps of 2.5  from  60 - 80, azimuth in steps of 10 from 0- 360 would require more than 5 million simulations. This means that the here presented ESRF exploration of the parameter space is more than 5000 times more efficient than a comparable parameter sweep. This effect increases if further free parameters are introduced.

\section{Conclusions}

    This study introduced data assimilation as a method to derive thermophysical model parameters along with their associated uncertainties from thermal infrared observations. The considered ESRF allows for a simultaneous estimation of the state, i.e., surface and subsurface temperatures, as well as model parameters, i.e. thermal inertia, emissivity, surface orientation etc., based on observed thermal infrared flux.
    Ensembles generated by the ESRF form a distribution that represents the uncertainties of state and parameters, while automatically including their respective correlations. As the performed forecast step is done on the basis of the thermal model without a linearisation the ensemble is able to better capture the nonlinear relations better than commonly employed techniques such as the Least squares method. 
    
    The observations of the MARA instrument onboard the MASCOT lander were revisited in this work applying the ESRF. The results are consistent with the initial analysis of \cite{grott2019} but narrow the range of the thermal inertia estimate to $295 \pm 18$ $\mathrm{J\,m^{-2}\,K^{-1}\,s^{-1/2}}$. At the same time the emissivity could be constrained to $0.95\pm0.05$. The average surface orientation of around $81^\circ \pm2^\circ$ elevation and $304^\circ \pm68^\circ$ azimuth indicate that a significant fraction of the boulder in the MARA field of view is orientated towards the instrument. As these parts of the boulder face away from the sun during day, this result is consistent with reduced daytime temperature and the roughness effect reported in \cite{grott2019}.
    
    In the first analysis of the MARA data, thermal conductivity $k$ and porosity $\phi$ of the boulder on Ryugu was estimated based on two empirical relations of $k(\phi)$ \cite{Henke2016,Flynn2018}. Repeating the calculation \cite{grott2019} for the thermal inertia estimate of this study, $295 \pm 18$ $\mathrm{J\,m^{-2}\,K^{-1}\,s^{-1/2}}$, results in $k = 0.11 \pm 0.01$ $\mathrm{W\,m^{-1}\,K^{-1}}$ and $\phi = 0.50 \pm 0.02$ using \cite{Flynn2018}, and $k = 0.08 \pm 0.01$ $\mathrm{W\,m^{-1}\,K^{-1}}$ and $\phi = 0.32 \pm 0.02$ using \cite{Henke2016}.
    
    An advantage of the ESRF scheme is its computational design to cope with large dimensions of state and parameter spaces, while being robust for nonlinear systems. The possibility of estimating many parameters simultaneously enhances the scientific output of the remote sensing data. The parameters retrieved from the MARA observations are more accurate than previous estimates, as the ESRF discards unlikely parameter combinations and incorporates their correlations. At the same time, the parameters were sampled from a wide section of the parameter space and allowed to vary freely, limited only by basic physical limits, which should ensure that the parameter space was sufficiently covered.
    
    The features of the family of EnKFs, i.e., coping with large dimensions of state and parameter spaces, provide great flexibility. The efficiency in handling many model parameters simultaneously sets the ESRF method apart from other Bayesian Monte-Carlo methods such as the Markov-Chain Monte Carlo \cite{cambioni2019} or Particle filters \cite[also ABC methods, see][]{ogawa2019}. Further applications may include other thermal models and sets of parameters, e.g. such as modelling multiple ground layers with different thermal conductivity, or including temperature dependent models of thermal conductivity and heat capacity \cite{Vasavada2017}. This will allow for an improved estimation of thermal properties on many other objects in the solar system such as Mars, the Moon, or Comets.

\section*{Code and Data Availability}
The code and data used in this study will be made available by the corresponding authors upon request.

\section*{Acknowledgements}
MH was financially supported by Geo.X, the Research Network for Geosciences in Berlin and Potsdam - SO\_087\_GeoX. The research of JdW and IP has been partially funded by Deutsche Forschungsgemeinschaft (DFG) - SFB1294/1 - 318763901.  JdW was also supported by ERC Advanced Grant “ACRCC” (grant 339390) and by the Simons CRM Scholar-in-Residence Program. We thank Dr. Andrew Ryan for his constructive and very helpful review.

\bibliographystyle{plain}
\bibliography{references}

\begin{thebibliography}{10}

\bibitem{cambioni2019}
Saverio Cambioni, Marco Delbo, Andrew~J Ryan, Roberto Furfaro, and Erik
  Asphaug.
\newblock Constraining the thermal properties of planetary surfaces using
  machine learning: Application to airless bodies.
\newblock {\em Icarus}, 325:16--30, 2019.

\bibitem{Chase1969}
S.C. Chase, Jr.
\newblock Infrared radiometer for the 1969 mariner mission to mars.
\newblock {\em Applied Optics}, 8:639--642, 1969.

\bibitem{christensen2018}
Philip~R Christensen, Victoria~E Hamilton, GL~Mehall, Daniel Pelham, William
  O’Donnell, Saadat Anwar, Heather Bowles, Stillman Chase, J~Fahlgren,
  Z~Farkas, et~al.
\newblock The osiris-rex thermal emission spectrometer (otes) instrument.
\newblock {\em Space Science Reviews}, 214(5):87, 2018.

\bibitem{Christensen2001}
P.R. Christensen, J.L. Bandfield, V.E. Hamilton, S.W. Ruff, H.H. Kieffer, T.N.
  Titus, M.C. Malin, R.V. Morris, M.D. Lane, R.L. Clark, B.M. Jakosky, M.T.
  Mellon, J.C. Pearl, B.J. Conrath, M.D. Smith, R.T. Clancy, R.O. Kuzmin,
  T.~Roush, N.~Mehall, G.L.and~Gorelick, K.~Bender, K.~Murray, S.~Dason,
  E.~Greene, S.~Silverman, and M.~Greenfield.
\newblock Mars global surveyor thermal emission spectrometer experiment:
  Investigation description and surface science results.
\newblock {\em J. Geophys. Res.}, 106(E10):23823--23871, 2001.

\bibitem{deWiljesStannatReich2019}
J.~de~Wiljes, W.~Stannat, and S.~Reich.
\newblock Long-time stability and accuracy of the ensemble kalman-bucy filter
  for fully observed processes and small measurement noise.
\newblock {\em SIAM J. Appl. Dyn. Syst}, 17(2):1152--1181, 2019.

\bibitem{deWiljesTong2019}
J.~de~Wiljes and X.~Tong.
\newblock Analysis of a localised nonlinear ensemble kalman bucy filter with
  complete and accurate observations.
\newblock {\em https://arxiv.org/abs/1908.10580}, 2019.

\bibitem{dellagiustina2019}
DN~DellaGiustina, JP~Emery, DR~Golish, Benjamin Rozitis, CA~Bennett, KN~Burke,
  R-L Ballouz, KJ~Becker, PR~Christensen, CY~Drouet d’Aubigny, et~al.
\newblock Properties of rubble-pile asteroid (101955) bennu from osiris-rex
  imaging and thermal analysis.
\newblock {\em Nature Astronomy}, 3(4):341, 2019.

\bibitem{Evensen2003}
G.~Evensen.
\newblock The ensemble kalman filter: Theoretical formulation and practical
  implementation.
\newblock {\em Ocean Dynamics}, 53:343–367, 2003.

\bibitem{jdw:EvensenLeeuwen2000}
G.~Evensen and P.~J. van Leeuwen.
\newblock An ensemble {K}alman smoother for nonlinear dynamics.
\newblock {\em Mon. Wea. Rev.}, 128(6):1852--1867, 2000.

\bibitem{evensen2006}
Geir Evensen.
\newblock {\em Data assimilation: the ensemble Kalman filter}.
\newblock Springer, 2006.

\bibitem{Fergason2006b}
R.~L. Fergason, P.~R. Christensen, J.~F. Bell, III, M.~P. Golombeck, K.~E.
  Herkenhoff, and H.~H. Kieffer.
\newblock Physical properties of the mars exploration rover landing sites as
  inferred from mini-tes-derived thermal inertia.
\newblock {\em Journal of Geophysical Research}, 111(E02S21), 2006.

\bibitem{Fergason2006a}
R.~L. Fergason, P.~R. Christensen, and H.~H. Kieffer.
\newblock High-resolution thermal inertia derived from the thermal emission
  imaging system (themis): Thermal model and applications.
\newblock {\em Journal of Geophysical Research}, 111(E12004), 2006.

\bibitem{Flynn2018}
George~J. Flynn, Guy~J. Consolmagno, Peter Brown, and Robert~J. Macke.
\newblock Physical properties of the stone meteorites: Implications for the
  properties of their parent bodies.
\newblock {\em Geochemistry}, 78(3):269 -- 298, 2018.

\bibitem{GomezElvira2012}
J.~{G{\'o}mez-Elvira}, C.~{Armiens}, L.~{Casta{\~n}er}, M.~{Dom{\'{\i}}nguez},
  M.~{Genzer}, F.~{G{\'o}mez}, R.~{Haberle}, A.-M. {Harri}, V.~{Jim{\'e}nez},
  H.~{Kahanp{\"a}{\"a}}, L.~{Kowalski}, A.~{Lepinette}, J.~{Mart{\'{\i}}n},
  J.~{Mart{\'{\i}}nez-Fr{\'{\i}}as}, I.~{McEwan}, L.~{Mora}, J.~{Moreno},
  S.~{Navarro}, M.~A. {de Pablo}, V.~{Peinado}, A.~{Pe{\~n}a}, J.~{Polkko},
  M.~{Ramos}, N.~O. {Renno}, J.~{Ricart}, M.~{Richardson},
  J.~{Rodr{\'{\i}}guez-Manfredi}, J.~{Romeral}, E.~{Sebasti{\'a}n},
  J.~{Serrano}, M.~{de la Torre Ju{\'a}rez}, J.~{Torres}, F.~{Torrero},
  R.~{Urqu{\'{\i}}}, L.~{V{\'a}zquez}, T.~{Velasco}, J.~{Verdasca}, M.-P.
  {Zorzano}, and J.~{Mart{\'{\i}}n-Torres}.
\newblock {REMS: The Environmental Sensor Suite for the Mars Science Laboratory
  Rover}.
\newblock {\em Space Science Review}, 170:583--640, 2012.

\bibitem{grott2017}
M~Grott, J~Knollenberg, B~Borgs, F~H{\"a}nschke, E~Kessler, J~Helbert,
  A~Maturilli, and N~M{\"u}ller.
\newblock The mascot radiometer mara for the hayabusa 2 mission.
\newblock {\em Space Science Reviews}, 208(1-4):413--431, 2017.

\bibitem{grott2019}
Matthias Grott, Joerg Knollenberg, Maximilian Hamm, Kazunori Ogawa, Ralf
  Jaumann, KA~Otto, Marco Delbo, Patrick Michel, Jens Biele, Wladimir Neumann,
  et~al.
\newblock Low thermal conductivity boulder with high porosity identified on
  c-type asteroid (162173) ryugu.
\newblock {\em Nature Astronomy}, pages 1--6, 2019.

\bibitem{Hamilton2014}
V.~E. Hamilton, A.~R. Vasavada, E.~Sebastian, M.~de~la Torre~Ju\'arez,
  M.~Ramos, C.~Armiens, R.~E. Arvidson, I.~Carrasco, P.~R. Christensen, M.~A.
  De~Pablo, W.~Goetz, J.~G\'omez-Elvira, M.~T. Lemmon, M.~B. Madsen, F.~J.
  Mart\'in-Torres, J.~Mart\'inez-Fr\'ias, A.~Molina, M.~C. Palucis, S.~C.~R.
  Rafkin, M.~I. Richardson, R.~A. Yingst, and M.~Zorzano.
\newblock Observations and preliminary science results from the first 100 sols
  of msl rover environmental monitoring station ground temperature sensor
  measurements at gale crater.
\newblock {\em Journal of Geophysical Research: Planets}, 119(4):745--770,
  2014.

\bibitem{hamm2018}
M.~Hamm, M.~Grott, E.~K\"uhrt, I.~Pelivan, and J.~Knollenberg.
\newblock A method to derive surface thermophysical properties of asteroid
  (162173) ryugu (1999ju3) from in-situ surface brightness temperature
  measurements.
\newblock {\em Planetary and Space Science}, 159:1--10, 2018.

\bibitem{Harris2016}
A.~W. Harris and L.~Drube.
\newblock Thermal tomography of asteroid surface structure.
\newblock {\em The Astrophysical Journal}, 827(2):127, 2016.

\bibitem{Henke2016}
Stephan Henke, Hans-Peter Gail, and Mario Trieloff.
\newblock Thermal evolution and sintering of chondritic planetesimals-iii.
  modelling the heat conductivity of porous chondrite material.
\newblock {\em Astronomy \& Astrophysics}, 589:A41, 2016.

\bibitem{ho2017}
Tra-Mi Ho, Volodymyr Baturkin, Christian Grimm, Jan~Thimo Grundmann, Catherin
  Hobbie, Eugen Ksenik, Caroline Lange, Kaname Sasaki, Markus Schlotterer,
  Maria Talapina, et~al.
\newblock Mascot—the mobile asteroid surface scout onboard the hayabusa2
  mission.
\newblock {\em Space Science Reviews}, 208(1-4):339--374, 2017.

\bibitem{hu2003adaptive}
Congwei Hu, Wu~Chen, Yongqi Chen, Dajie Liu, et~al.
\newblock Adaptive kalman filtering for vehicle navigation.
\newblock {\em Journal of Global Positioning Systems}, 2(1):42--47, 2003.

\bibitem{jaumann2019}
R~Jaumann, N~Schmitz, T-M Ho, SE~Schr{\"o}der, KA~Otto, K~Stephan, S~Elgner,
  K~Krohn, F~Preusker, F~Scholten, et~al.
\newblock Images from the surface of asteroid ryugu show rocks similar to
  carbonaceous chondrite meteorites.
\newblock {\em Science}, 365(6455):817--820, 2019.

\bibitem{jdw:Kalman1960}
R.~E. Kalman.
\newblock A new approach to linear filtering and prediction problems.
\newblock {\em Transaction of the ASME Journal of Basic Engineering}, pages
  35--45, 1960.

\bibitem{Kieffer1972}
Hugh~H. Kieffer, G.~Neugebauer, G.~Munch, S.C. Chase, and E.~Miner.
\newblock Infrared thermal mapping experiment: The viking mars orbiter.
\newblock {\em Icarus}, 16(1):47 -- 56, 1972.

\bibitem{Kuehrt1992}
E.~K\"uhrt, B.~Giese, H.~U. Keller, and L.V Ksanfomality.
\newblock Interpretation of the krfm-infrared measurements of phobos.
\newblock {\em Icarus}, 96(2):213--218, 1992.

\bibitem{LangeStannat2019}
T.~Lange and W.~Stannat.
\newblock On the continuous time limit of ensemble square root filters.
\newblock {\em https://arxiv.org/abs/1910.12493}, 24:118--173, 2019.

\bibitem{lauretta2019}
DS~Lauretta, DN~DellaGiustina, CA~Bennett, DR~Golish, KJ~Becker,
  SS~Balram-Knutson, OS~Barnouin, TL~Becker, WF~Bottke, WV~Boynton, et~al.
\newblock The unexpected surface of asteroid (101955) bennu.
\newblock {\em Nature}, 568(7750):55, 2019.

\bibitem{jdw:stuart15}
K.~Law, A.~Stuart, and K.~Zygalakis.
\newblock {\em Data Assimilation: A Mathematical Introduction}.
\newblock Springer-Verlag, New York, 2015.

\bibitem{Masiero2011}
Joseph~R Masiero, AK~Mainzer, T~Grav, JM~Bauer, RM~Cutri, J~Dailey, PRM
  Eisenhardt, RS~McMillan, TB~Spahr, MF~Skrutskie, et~al.
\newblock Main belt asteroids with wise/neowise. i. preliminary albedos and
  diameters.
\newblock {\em The Astrophysical Journal}, 741(2):68, 2011.

\bibitem{Mellon2000}
M.T. Mellon, B.M. Jakosky, H.H. Kieffer, and P.R Christensen.
\newblock High-resolution thermal inertia mapping from the mars global surveyor
  thermal emission spectrometer.
\newblock {\em Icarus}, 148(2):437--455, 2000.

\bibitem{2006Icar..185..113M}
L.~{Montabone}, S.~R. {Lewis}, P.~L. {Read}, and D.~P. {Hinson}.
\newblock {Validation of martian meteorological data assimilation for MGS/TES
  using radio occultation measurements}.
\newblock {\em Icarus}, 185:113--132, November 2006.

\bibitem{mueller2017}
T.~G. M\"uller, J.~Durech, M.~Ishiguro, M.~Mueller, T.~Kr\"uhler, H.~Yang,
  M.~J. Kim, L.~O’Rourke, F.~Usui, C.~Kiss, B.~Altieri, B.~Carry, Y.~J. Choi,
  M.~Delbo, J.~P. Emery, J.~Greiner, S.~Hasegawa, J.~L. Hora, F.~Knust,
  D.~Kuroda, D.~Osip, A.~Rau, A.~Rivkin, P.~Schady, J.~Thomas-Osip,
  D.~Trilling, S.~Urakawa, E.~Vilenius, P.~Weissman, and P.~Zeidler.
\newblock Hayabusa-2 mission target asteroid 162173 ryugu (1999 ju3): Searching
  for the object’s spin-axis orientation.
\newblock {\em Astronomy \& Astrophysics}, 599:A103, 2017.

\bibitem{mueller2014}
T.~G. M\"uller, S.~Hasegawa, and F.~Usui.
\newblock (25143) itokawa: The power of radiometric techniques for the
  interpretation of remote thermal observations in the light of the hayabusa
  rendezvous results.
\newblock {\em Publications of the Astronomical Society of Japan},
  66(3):52--52, 2014.

\bibitem{Nergeretal2012}
L.~Nerger, T.~Janjić, J.~Schroeter, and W.~Hiller.
\newblock A unification of ensemble square root filters.
\newblock {\em Monthly Weather Review}, 140:2335--2345, 2012.

\bibitem{Nowicki2007}
S.~A. Nowicki and P.~R. Christensen.
\newblock Rock abundance on mars from the thermal emission spectrometer.
\newblock {\em Journal of Geophysical Research: Planets}, 112(E5), 2007.

\bibitem{okada2020}
Tatsuaki Okada, Tetsuya Fukuhara, Satoshi Tanaka, Makoto Taguchi, Takehiko
  Arai, Hiroki Senshu, Naoya Sakatani, Yuri Shimaki, Hirohide Demura, Yoshiko
  Ogawa, et~al.
\newblock Highly porous nature of a primitive asteroid revealed by thermal
  imaging.
\newblock {\em Nature}, pages 1--5, 2020.

\bibitem{okada2017}
Tatsuaki Okada, Tetsuya Fukuhara, Satoshi Tanaka, Makoto Taguchi, Takeshi
  Imamura, Takehiko Arai, Hiroki Senshu, Yoshiko Ogawa, Hirohide Demura, Kohei
  Kitazato, et~al.
\newblock Thermal infrared imaging experiments of c-type asteroid 162173 ryugu
  on hayabusa2.
\newblock {\em Space Science Reviews}, 208(1-4):255--286, 2017.

\bibitem{Paige2010}
D.~A. Paige, M.~C. Foote, B.~T. Greenhagen, J.~T. Schofield, S.~Calcutt, A.~R.
  Vasavada, D.~J. Preston, F.~W. Taylor, C.~C. Allen, K.~J. Snook, B.~M.
  Jakosky, B.~C. Murray, L.~A. Soderblom, B.~Jau, S.~Loring, J.~Bulharowski,
  N.~E. Bowles, I.~R. Thomas, M.~T. Sullivan, C.~Avis, E.~M. De~Jong,
  W.~Hartford, and D.~J. McCleese.
\newblock The lunar reconnaissance orbiter diviner lunar radiometer experiment.
\newblock {\em Space Science Reviews}, 150(1):125--160, 2010.

\bibitem{Pelivan2017}
I.~Pelivan, L.~Drube, E.~K{\"u}hrt, J.~Helbert, J.~Biele, M.~Maibaum,
  B.~Cozzoni, and V.~Lommatsch.
\newblock Thermophysical modeling of didymos' moon for the asteroid impact
  mission.
\newblock {\em Advances in Space Research}, 59(7):1936 -- 1949, 2017.

\bibitem{preusker2019}
Frank Preusker, Frank Scholten, Stephan Elgner, K-D Matz, Shingo Kameda, Thomas
  Roatsch, R~Jaumann, S~Sugita, R~Honda, T~Morota, et~al.
\newblock The mascot landing area on asteroid (162173) ryugu:
  Stereo-photogrammetric analysis using images of the onc onboard the hayabusa2
  spacecraft.
\newblock {\em Astronomy \& Astrophysics}, 632:L4, 2019.

\bibitem{jdw:reichcotter15}
S.~Reich and C.J. Cotter.
\newblock {\em Probabilistic Forecasting and Bayesian Data Assimilation}.
\newblock Cambridge University Press, Cambridge, 2015.

\bibitem{sakatani2017}
N.~Sakatani, K.~Ogawa, Y.~Iijima, M.~Arakawa, R.~Honda, and S.~Tanaka.
\newblock Thermal conductivity model for powdered materials under vacuum based
  on experimental studies.
\newblock {\em AIP Advances}, 7, 2017.

\bibitem{scholten2019descent}
Frank Scholten, Frank Preusker, Stephan Elgner, K-D Matz, R~Jaumann, Jens
  Biele, D~Hercik, H-U Auster, Maximilian Hamm, Matthias Grott, et~al.
\newblock The descent and bouncing path of the hayabusa2 lander mascot at
  asteroid (162173) ryugu.
\newblock {\em Astronomy \& Astrophysics}, 632:L3, 2019.

\bibitem{scholten2019boulder}
Frank Scholten, Frank Preusker, Stephan Elgner, K-D Matz, R~Jaumann, Maximilian
  Hamm, SE~Schr{\"o}der, Alexander Koncz, Nicole Schmitz, Frank Trauthan,
  et~al.
\newblock The hayabusa2 lander mascot on the surface of asteroid (162173)
  ryugu--stereo-photogrammetric analysis of mascam image data.
\newblock {\em Astronomy \& Astrophysics}, 632:L5, 2019.

\bibitem{Spohn2015}
T.~Spohn, J.~Knollenberg, A.~J. Ball, M.~Banaszkiewicz, J.~Benkhoff, M.~Grott,
  J.~Grygorczuk, C.~Huttig, A.~Hagermann, G.~Kargl, E.~Kaufmann, N.~Komle,
  E.~Kuhrt, K.~J. Kossacki, W.~Marczewski, I.~Pelivan, R.~Schrodter, and
  K.~Seiferlin.
\newblock Thermal and mechanical properties of the near-surface layers of comet
  67p/churyumov-gerasimenko.
\newblock {\em Science}, 349(6247):aab0464, 2015.

\bibitem{sugita2019}
Satoshi Sugita, Rie Honda, Tomokatsu Morota, Shingo Kameda, Hirotaka Sawada,
  Eisuke Tatsumi, Matsuichi Yamada, Chikatoshi Honda, Yasuhiro Yokota, Toru
  Kouyama, et~al.
\newblock The geomorphology, color, and thermal properties of ryugu:
  Implications for parent-body processes.
\newblock {\em Science}, 364(6437):eaaw0422, 2019.

\bibitem{jdw:tippett03}
M.K. Tippett, J.L. Anderson, G.H. Bishop, T.M. Hamill, and J.S. Whitaker.
\newblock Ensemble square root filters.
\newblock {\em Mon. Wea. Rev.}, 131:1485--1490, 2003.

\bibitem{Vasavada2017}
A.~R. Vasavada, S.~Piqueux, K.~W. Lewis, M.~T. Lemmon, and M.~D. Smith.
\newblock Thermophysical properties along curiosity 's traverse in gale crater,
  mars, derived from the rems ground temperature sensor.
\newblock {\em Icarus}, 284:372--386, 2017.

\bibitem{wada2018}
Koji Wada, Matthias Grott, Patrick Michel, Kevin~J Walsh, Antonella~M Barucci,
  Jens Biele, J{\"u}rgen Blum, Carolyn~M Ernst, Jan~Thimo Grundmann, Bastian
  Gundlach, et~al.
\newblock Asteroid ryugu before the hayabusa2 encounter.
\newblock {\em Progress in Earth and Planetary Science}, 5(1):82, 2018.

\bibitem{watanabe2017}
Sei-ichiro Watanabe, Yuichi Tsuda, Makoto Yoshikawa, Satoshi Tanaka, Takanao
  Saiki, and Satoru Nakazawa.
\newblock Hayabusa2 mission overview.
\newblock {\em Space Science Reviews}, 208(1-4):3--16, 2017.

\bibitem{watanabe2019}
Seiichiro Watanabe, M~Hirabayashi, N~Hirata, Na~Hirata, R~Noguchi, Y~Shimaki,
  H~Ikeda, E~Tatsumi, M~Yoshikawa, S~Kikuchi, et~al.
\newblock Hayabusa2 arrives at the carbonaceous asteroid 162173 ryugu—a
  spinning top--shaped rubble pile.
\newblock {\em Science}, 364(6437):268--272, 2019.

\bibitem{2008GeoRL..35.7202W}
R.~J. {Wilson}, S.~R. {Lewis}, L.~{Montabone}, and M.~D. {Smith}.
\newblock {Influence of water ice clouds on Martian tropical atmospheric
  temperatures}.
\newblock {\em Geophysical Research Letters}, 35:L07202, April 2008.

\end{thebibliography}
\end{document}